\documentstyle[aps,preprint, eqsecnum]{revtex}
\begin{document}
\draft
\tightenlines

\title{ A boost to low-energy nuclear reactions
from preliminary elastic collisions\\
(Carambole collisions) }

\author{M.Yu. Kuchiev$^{1}$ \thanks{email address: 
kuchiev@newt.phys.unsw.edu.au},
B.L. Altshuler$^{2}$, V.V. Flambaum$^{1}$ }

\address{$^1$ School of Physics, University of New South Wales,
Sydney 2052, Australia}

\address{
$^2$ Physics Department, Princeton University, Princeton, NJ 08544
 and NEC Research 
Institute, 4 Independence Way, Princeton NJ 08540, USA}

\date{\today}
\maketitle


\begin{abstract}
In conventional nuclear
experiments a beam of accelerated nuclei collides with a target
nucleus that is surrounded by other  nuclei
in a molecule, in condensed matter, or in a plasma environment.
It is shown that for 
low collision energies 
possible nuclear reactions (including deuterium fusion)
are
strongly boosted by the environment.  The effect
originates from a chain of  preliminary
{\em three  elastic collisions}
which transform the projectile-target
experiment into one with colliding beams.
Firstly, the projectile-target pair of
nuclei undergo elastic scattering 
in which the
projectile shares its energy and momentum with the target nucleus.
Then the projectile and target nuclei collide with different 
heavy nuclei from the environment.
These latter collisions change the velocities of the target and
projectile nuclei and set them again on the collision course.
Finally, the same pair of nuclei collide inelastically,
this time giving rise to the nuclear reaction.
The increased energy of the target
nucleus increases the relative velocity up to $\sqrt 2$ times,
resulting in a drastic exponential increase of the
probability to penetrate the Coulomb barrier and therefore
sharply increasing the likelihood of the nuclear reaction.
Applications to laser-induced fusion are discussed.
\end{abstract}

\pacs{PACS numbers: 34.50-s, 34.90.+q, 25.60.Pj }


\section{Introduction}
\label{intro}

We suggest a new mechanism
that strongly increases
the probability of low-energy nuclear reactions in condensed matter,
in molecules, or in a plasma environment.
Although we do not aim to interpret a particular experiment, 
our interest in the subject was definitely inspired by
a few publications, see  \cite{1,2,Pd,Ichimaru}
and references therein,
that claim  a substantial increase
of the DD fusion cross-section in solids. 
One obvious possibility for a boost of the fusion 
originates from the motion
of the target deuterium caused by its vibrations
in a condensed matter (or other) environment, as discussed
in \cite{AFKZ}. Moreover, the vibration can be stimulated by
the Coulomb interaction of the projectile
deuterium with the target one, giving an
additional increase for the probability of the fusion,
as was found in \cite{AFKZ}. In this paper
we suggest another effect which proves to be
very effective in the low-energy region.

Consider some nuclear reaction initiated by the collision 
of a pair of nuclei in a conventional beam-type nuclear experiment. 
Suppose that the target nucleus is
surrounded by other heavy nuclei.
This happens when it is deposited in a molecule,
in condensed matter, or in a plasma environment.
The energy of the projectile is supposed to be lower than
the Coulomb barrier, and we are interested in the
probability of a nuclear reaction.
Remember that the Coulomb barrier
makes the probability $P$ of the reaction depend exponentially
on the relative velocity $v_{12}$ of the colliding pair \cite{Landau}
\begin{equation}\label{prob}
P \propto \exp \left(- \frac{2 \pi e^2 Z_1 Z_2}{\hbar v_{12} } \right)~,
\end{equation}
where $Z_1,Z_2$ are the charges of the projectile
and target nuclei, $v_{12}$ is their relative velocity.
(Here and afterwards the indexes $1$ and $2$ mark the 
velocities, momenta, and energies
of the projectile and target nuclei, respectively.)
If we assume that the energy of the target nucleus $E_2$ is negligible,
$E_2 \sim 0$, then
the collision velocity depends only on the energy of the projectile 
$E_1,~v_{12} \simeq v_1 = \sqrt{ 2 E_1/m_1}$, making
the probability (\ref{prob}) 
depend exponentially on this energy.
Let us imagine now that
there  exists some mechanism 
which can split the same amount of energy between 
the two colliding nuclei. Moreover, let us
assume that this mechanism can  also 
put the nuclei in a head on collision.
Then obviously the relative velocity may be larger. Its
possible maximum for a given energy $E_1$ is $v_m =\sqrt{ 2 E_1/\mu}$,
where $\mu$ is the reduced mass for the pair of nuclei.
For the collision of two deuterons
$m_1=m_2 =2\mu$, 
then the maximum possible  relative velocity $v_m = \sqrt{2}\, v_1$ is 
substantially larger than the initial velocity, $v_m > v_1$. 
This increase of the relative velocity and a corresponding gain in the energy 
of the relative motion is due to reduction 
of the energy of the center of mass 
motion which was a half of the total energy in the initial projectile
deuteron problem.
Eq.(\ref{prob}) shows that collision with
the velocity $v_m= \sqrt 2 \, v_1$ 
makes the nuclear reaction much more probable.

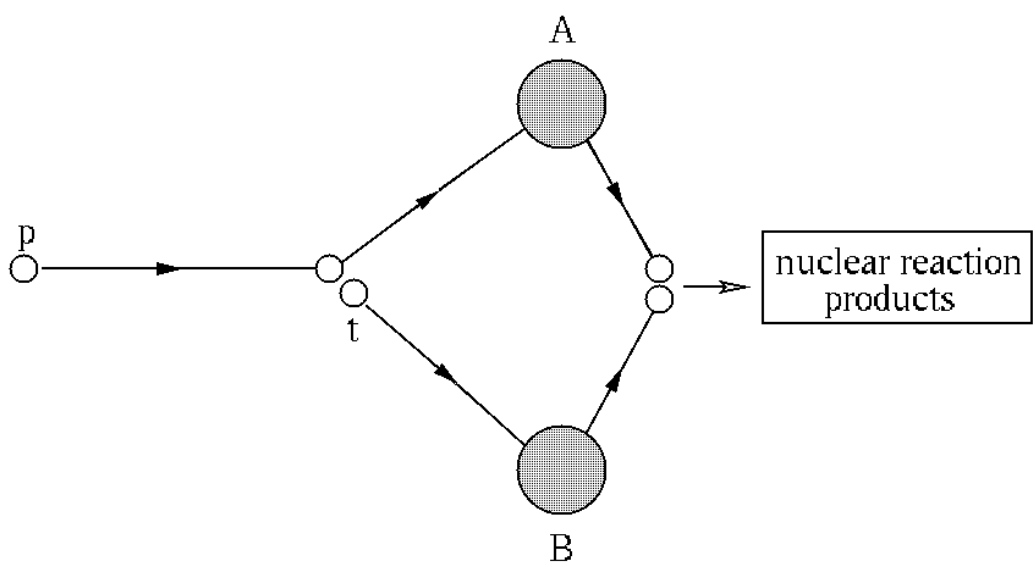
\begin{figure}
\caption{ \label{cartoon}
The sketch illustrates the main idea of three elastic collisions.
Lines with arrows show classical trajectories of 
the projectile and target nuclei that are depicted by small circles.
The projectile nucleus collides firstly
with the target nucleus,
then the projectile and target nuclei are scattered by
atoms A and B depicted by large  circles. After that
the inelastic collision 
of the projectile and target nuclei takes place,
resulting in the nuclear reaction. 
The relative velocity in this 
latter collision is larger than the
initial velocity of the projectile nucleus, which
enhances the probability of the event.}
\end{figure}

This simple observation inspires one to look for a mechanism which
can fulfill two functions. 
Firstly, it should force the projectile
to share its energy with the target. Secondly, 
it has to set the pair onto the collision course.
The first goal can be achieved if we
consider an {\em elastic} scattering of the projectile on the target 
nucleus.
This collision  obviously transfers part of the
projectile energy to the target and can
substantially enlarge the
relative velocity of the two nuclei.
This event should be considered as a {\em preliminary} collision, 
since we are seeking the possibility for the nuclear
reaction to occur in the {\em following}
inelastic collision.
However, the necessary inelastic event cannot happen by itself.
After the elastic collision, the two nuclei move
in opposite directions in the center-of-mass reference frame,
and therefore cannot collide again, at least 
this is impossible in the vacuum. 
The situation changes when the environment is present. 
The important point is that
scattering of the considered colliding  particles on
nuclei of the environment
can change the directions of their propagation.
To simplify the consideration, let us assume that
the nuclei of the environment are much heavier than the 
colliding pair (which is usually true for the
most interesting case of light colliding nuclei).
Then, the scattering on nuclei of the environment
does not change the energies of the target and projectile nuclei,
but changes the directions of their velocities.
These new velocities may put
the pair on the collision course.
There arises therefore a possibility for 
the second collision of the same projectile-target pair.
A sketch in Fig.\ref{cartoon} illustrates this idea.
Since this final collision happens with larger 
relative
velocity,
it can initiate the nuclear reaction with much higher probability.

The picture described  needs a chain of {\em three
elastic collisions} (TEC).
The initial elastic collision of the pair
accelerates the target, then the two collisions of the target
and projectile nuclei with different nuclei of the environment
change the relative velocity of the pair in such a way that the
final inelastic collision becomes possible.
One can find some resemblance of this process with the 
cannon (carambole) shot in the billiards game.

The collisions of the projectile and target nuclei with
atoms of the environment can, generally speaking,
result in ionization of these atoms. 
One can expect, however, that an energy loss due to ionization
is much lower than the collision energy, provided
the latter is higher than the atomic ionization potential.
That is why even if the ionization happens, the nuclei can still
move along the trajectories which lead to their final inelastic
collision. In other words, it is 
sufficient for our purposes
to assume that collisions with atoms A and B 
happen quasielastically. The colliding nuclei
do not lose a substantial amount of their energy, but
transfer a large momentum.

Proposing the TEC mechanism,
we find  that it greatly, by orders of magnitude, 
increases the probability of the nuclear 
reaction for a low projectile energy. 
In Section \ref{wf} we find the wave function for the
colliding pair which takes into account TEC. 
In Sections \ref{amplit} and \ref{proba}
we derive the amplitude and the probability of the nuclear reaction
initiated by TEC.
Section \ref{screen} examines the role of screening, while
in Section \ref{num} we present a numerical example which shows how 
effective the TEC mechanism is for deuterium fusion.
The maximum effect is achieved when the target deuterium is located
at the center of the square formed by heavy atoms 
and the beam 
of projectile deuterons is parallel to the
surface of this square.
Conclusions are presented in Section \ref{con}.

\section{Wave function}
\label{wf}
Our goal is to find the wave function which takes into account TEC.
Note firstly that each step of TEC
happens in some region, and the higher the collision energy, the 
more localized is the region.
After each collision the active nucleus
propagates a long distance before colliding again.
Fig.\ref{cartoon} illustrates this statement.
This indicates that each of the TEC collisions may be described
in terms of the corresponding scattering amplitudes (physical amplitudes,
i.e. {\em on mass shell} amplitudes).
Let us call $f_{pt}(\theta_{pt})$
the scattering amplitude for
the elastic collision of the projectile with the target nucleus, 
and $f_{pA}(\theta_{pA}),~ f_{tB}(\theta_{tB})$ 
the scattering amplitudes
for collisions of the projectile nucleus with an atom $A$ and the target
nucleus
with an atom $B$. We expect that
the wave function of the target-projectile pair of nuclei which takes
TEC into account
$\delta ^{(TEC)} \psi({\bf r}_p,{\bf r}_t)$ should be proportional
to these amplitudes, 
$\delta ^{(TEC)} \psi({\bf r}_p,{\bf r}_t) \propto 
f_{pt}(\theta_{pt})\, f_{pA}(\theta_{pA})\,f_{tB}(\theta_{tB})$.

Furthermore,  one can describe the
propagation of the target and projectile nuclei 
in between the collisions in terms of
the semiclassical approximation which is valid since
the nuclei are heavy in comparison with electrons.
Therefore we need to find the classical trajectories
which describe this propagation.
This many-body problem can be greatly simplified 
if we assume that
in between the elastic collisions
nuclei  propagate along straight lines.
This requires that the collision energy satisfies
\begin{equation}\label{qlarge}
\frac{p_1^2}{2m_1 } \gg Z_p \cdot Z_t \,\cdot \,27 ~{\rm eV}~,
\end{equation}
where $p_1, m_p$ are the momentum and mass of the projectile nucleus,
and $Z_p,Z_t$ are the charges of the projectile and target nuclei.
Condition (\ref{qlarge}) is valid for any reasonable
energy of the projectile. 
The relevant classical trajectories for both projectile and target nuclei
start at some initial point
${\bf r}_i$, where the first projectile-target elastic collision 
takes place.
From this point 
the projectile propagates with constant velocity ${\bf v}_1'$ to the point
${\bf r}_A$ where an atom A is located, and then, after scattering,
acquires the velocity ${\bf v}_1''$ and 
moves to the final point ${\bf r}_p$.
Similarly, the target moves from the point ${\bf r}_i$ 
with the velocity ${\bf v}_2'$
towards an atom 
B located at ${\bf r}_B$ and then with the velocity 
${\bf v}_2''$ to the final point
${\bf r}_t$. 
Fig. \ref{feynman_d} shows the Feynman diagram which
illustrates this discussion.
Obviously, the initial  projectile-target collision
preserves the total energy and momentum of the pair, while collisions
of the projectile and target nuclei
with atoms $A$ and $B$ preserve the absolute values of
their momenta, provided the atoms A and B are sufficiently heavy.
\footnote{Atoms A and B can be ionized during collisions.
This process, generally speaking, reduces the total energy of the pair.
However, this energy loss is comparable with the atomic ionization
potential and, consequently, is much lower than the total energy of the pair.
This allows us to neglect the ionization process 
in the following consideration.}
These classical conditions fix the classical trajectories
for the projectile and target nuclei and
allow one to express all relevant parameters, such as ${\bf r}_i~,
{\bf v}_p',~{\bf v}_t',~{\bf v}_p'',~{\bf v}_t''$  {\em etc},
as functions of the final coordinates ${\bf r}_p,~{\bf r}_t$
and initial momenta ${\bf p}_1,{\bf p}_2$ of the projectile and target nuclei.
From the classical trajectories  one can construct
the corresponding semiclassical wave function 
\begin{eqnarray}\label{FS}
&& \delta^{(TEC)}\psi_{ {\bf p}_1,{\bf p}_2 } ({\bf r},{\bf r}) =
F \exp \frac{i}{\hbar}S~,
\\ \label{Fxxx}
&&F = {\cal X} \, 
f_{pt}(\theta_{pt})\, f_{pA}(\theta_{pA})\,f_{tB}(\theta_{tB})~,
\end{eqnarray}
where $S$ is the classical action for the described trajectories,
and $F$ is the corresponding pre-exponential factor. Eq.~(\ref{Fxxx})
takes into account the fact that the wave function should be proportional
to the scattering amplitudes for TEC. From the simple dimensional
analysis we conclude that the only unknown parameter 
${\cal X}$ in (\ref{Fxxx})
should have the dimension of (length)$^{-3}$. The only available
parameter of length in the problem 
is a separation of atoms $a$ in the 
condensed matter (or molecule, or plasma) environment. Therefore
we find an estimate ${\cal X}\sim 1/a^3$, which agrees with
the more involved calculations discussed below.

Remember that we need the wave function in order to describe 
the final inelastic collision
of the pair. This requires that the projectile and target nuclei
eventually come to the same point ${\bf r}$, where the
final collision takes place. Therefore it is sufficient for us
to consider the wave function only at this point, i.e. when
${\bf r}_p={\bf r}_t={\bf r}$.
The method outlined above is discussed in detail
for this specific case in Appendix A.
The found wave function 
(\ref{psi3})  has the form
\begin{eqnarray}\label{sc}
\delta^{(TEC)}\psi_{ {\bf p}_1,{\bf p}_2 } ({\bf r},{\bf r}) =
\frac{f_{pt}(\theta_{pt})}{ R({\bf r}) }\,
\frac{  f_{pA}(\theta_{pA})}{ |{\bf r}-{\bf r}_A| }\, 
\frac{ f_{tB}(\theta_{tB})}
{ |{\bf r}- {\bf r}_B |}\, \exp \frac{i}{\hbar} S~, 
\end{eqnarray}
where the classical action $S$ is given by (\ref{phasa}).
The wave function (\ref{sc}) is proportional
to the scattering amplitudes, as was anticipated;
the corresponding scattering angles
$\theta_{pt},\theta_{pA},\theta_{tB}$
can be found from the classical trajectories.
The wave function is also inversely proportional to the factors 
$|{\bf r}-{\bf r}_A|,~|{\bf r}-{\bf r}_B| $, as it should be since
the waves that describe  
the projectile and target nuclei at the final point
${\bf r}$
originate from the points ${\bf r}_A$ and ${\bf r}_B$, 
respectively. The factor $R({\bf r})$ in the denominator is
an effective distance describing the propagation of the 
projectile
and target nuclei from their common initial point, 
 ${\bf r}_i$ to ${\bf r}_A$ and ${\bf r}_B$, respectively.
Its explicit form is presented in Eq.~(\ref{RS''}),
but for our purposes it is sufficient to
keep in mind that it is comparable with atomic separation 
\begin{equation}\label{rrab}
R({\bf r}) \simeq r_{AB}~.
\end{equation}
The semiclassical wave function  (\ref{sc}) agrees
with a simple estimate (\ref{FS}),~(\ref{Fxxx}).

Eq.~(\ref{sc}) is valid for the case
when both the projectile and
target nuclei in the initial state are described by
plane waves.  The validity of the plane wave approximation 
for the projectile nucleus is
justified by its sufficiently high energy
(\ref{qlarge}).
In contrast, the motion of the target nucleus 
arises  due to  
vibrations of this nucleus in a molecule or in condensed matter, or
by temperature reasons in a plasma environment.
Therefore, generally speaking, the wave function of the target nucleus
is not a plane wave. However,  typical target 
momenta are small compared with  the projectile momentum, 
\begin{equation}\label{p2<p1}
p_2 \ll p_1~.
\end{equation}
This fact allows one to  modify the approach outlined above to
accommodate the vibration motion of the target nucleus.
Suppose that the target nucleus is described by the wave function
$\phi_t({\bf r})$ in the initial state.
Consider  an
event in which the target nucleus possesses a momentum ${\bf p}_2$
before the collision.
The amplitude of this event equals the 
the Fourier component of the target wave function,  
$\tilde \phi_t({\bf p}_2)$.
Multiplying (\ref{sc}) by this amplitude, we find 
the contribution which this  event gives to the wave function
of the pair.
Integrating over all possible momenta of the target, we
obtain 
the following wave
function for the pair,
\begin{equation}\label{3sphi}
\delta^{(TEC)} \Psi ({\bf r}_1, {\bf r}_2)=
\int
\delta^{(TEC)}\psi_{ {\bf p}_1,{\bf p}_2 } ({\bf r},{\bf r}) 
\tilde \phi_t({\bf p}_2) \frac{d^3p_2}{(2\pi \hbar)^3}~,
\end{equation}
that  takes into account a distribution of the
target momentum in the initial state.
This approach, justified by inequality (\ref{p2<p1}), 
is similar to the momentum approximation
that is well known in nuclear physics \cite{Migdal}.

The semiclassical nature of the problem allows one to 
make the next step,
fulfilling integration in (\ref{3sphi}) explicitly.
Note that according to (\ref{p2<p1})
only small momenta $p_2$ contribute
to the integral. Therefore we can consider the limit
$p_2 \rightarrow 0$ in the wave function 
$\delta^{(TEC)}\psi_{ {\bf p}_1,{\bf p}_2 } ({\bf r},{\bf r})$.
In this limit the classical action $S({\bf r},{\bf p}_1,{\bf p}_2)$ 
should be expanded in the Taylor series
\begin{equation}\label{ser}
S({\bf r},{\bf p}_1,{\bf p}_2) = 
S({\bf r},{\bf p}_1,{\bf 0}) + {\bf r}_i \cdot {\bf p}_2
 + \dots ~,
\end{equation}
where  it is taken into account that the derivative
of the classical action
over the initial momentum gives the initial coordinate
for the considered trajectory,
\begin{equation}\label{fd}
\frac{ \partial S({\bf r},{\bf p}_1,{\bf p}_2) }{ \partial {\bf p}_2 }=
{\bf r}_i~.
\end{equation}
Higher derivatives, denoted by dots in (\ref{ser}), are negligible.
The action is the only quantity in the wave function
$\delta^{(TEC)}\psi_{ {\bf p}_1,{\bf p}_2 } ({\bf r},{\bf r})$
which needs to be considered accurately, using
the series expansion. For other parameters
one can simply take the limit  ${\bf p}_2 = {\bf 0}$.
As a result we derive from  (\ref{ser})
\begin{equation}\label{wf0}
\delta^{(TEC)}\Psi_{ {\bf p}_1,{\bf p}_2 } ({\bf r},{\bf r})=
\delta^{(TEC)}\psi_{ {\bf p}_1,{\bf p}_2= {\bf 0} } ({\bf r},{\bf r})
\exp \frac{i}{\hbar} ({\bf r}_i \cdot {\bf p}_2 )~.
\end{equation}
Substituting this into (\ref{3sphi})
we find the final  expression for the wave function
\begin{equation}\label{Psi}
\delta^{(TEC)} \Psi ({\bf r}, {\bf r})=
\delta^{(TEC)}\psi_{ {\bf p}_1,{\bf p}_2={\bf 0}} ({\bf r},{\bf r}) 
\,\phi_t({\bf r}_i)~,
\end{equation}
where $\phi_t({\bf r}_i)$ is the wave function of the target nucleus
calculated at the point ${\bf r}_i$, the latter one
is a function of ${\bf r}$ as is discussed above
(see also Appendix A).
Eq.~(\ref{Psi})
has a clear physical meaning. In order to set
the TEC mechanism in action, 
the initial collision of the projectile and
target nuclei should happen at some particular initial point
${\bf r}_i$, starting from which the
projectile and target propagate and undergo rescattering,
arriving eventually at the final point ${\bf r}$.
Therefore, the wave function of the pair should be proportional to the
amplitude to find the target located at the point ${\bf r}_i$.
This later amplitude
is simply equal to the wave function of the target nucleus
$\phi_t({\bf r}_i)$.
Since the momentum of the target is much smaller than that of the
projectile, we can 
neglect in (\ref{Psi}) the initial slow motion of the target
nucleus
\footnote{ The role of this motion is investigated
in some detail in  \cite{AFKZ}.},
and describe what happens after the  initial collision with
the help of the function
$\delta^{(TEC)}\psi_{ {\bf p}_1,{\bf p}_2={\bf 0}} ({\bf r},{\bf r})$.

Eqs. (\ref{Psi}), (\ref{sc})   are the main result of this section.
They present the wave function for the TEC problem in terms of classical
trajectories that describe the propagation of the projectile and
target nuclei between the collisions, the scattering amplitudes 
responsible for the elastic collisions, 
and the wave function of the target nucleus.

\section{Amplitude of nuclear reaction}
\label{amplit}

Let us call $A ( {\bf p}_1, {\bf p}_2 )$
the amplitude of the nuclear reaction initiated by a  collision
of  the pair of nuclei with momenta ${\bf p}_1$ and $ {\bf p}_2 $
in a vacuum.
According to (\ref{prob}) this amplitude strongly depends
on the collision velocity. It is convenient to account for
this dependence by introducing the amplitude
$B ( {\bf p}_1, {\bf p}_2 )$,
\begin{equation}\label{tun}
A ( {\bf p}_1, {\bf p}_2 ) = 
\exp \left( - \frac{  \pi e^2\,Z_1 Z_2 }{\hbar v}\right)
B ( {\bf p}_1, {\bf p}_2 )~.
\end{equation}
Here the exponential factor describes  
the under-barrier penetration through the Coulomb barrier
(that strongly depends on the velocity 
$v = |{\bf p}_1/m_p-{\bf p}_2/m_t|$),
while
$B ( {\bf p}_1, {\bf p}_2 )$ describes the proper 
nuclear processes which happen afterwards.

Let us express the amplitudes of events which happen
in the environment (in condensed matter, molecules, {\em etc})
in terms of the vacuum amplitude.
Consider first the {\em direct} collision of the projectile
with the target, when one neglects preliminary elastic rescattering.
In this case
the wave function of the pair 
is simply a product of the
incident plane wave describing the projectile
nucleus and the target wave function,
\begin{equation} \label{prod}
\psi_{{\bf p}_1 }( {\bf r}_p,{\bf r}_t) = 
\exp \frac{i}{\hbar}  ({\bf p}_p \cdot {\bf r}_p)\,
\phi_t({\bf r}_t)~.
\end{equation}
The amplitude of the {\em direct} 
nuclear reaction, in which the reaction products
in the final state
have momentum ${\bf Q}$, can be written as
\begin{eqnarray}\label{M}
M^{(DIR)} ( {\bf Q} ) &=& 
A( {\bf p}_1, {\bf Q}-{\bf p}_1 ) \,
\int  \psi_{{\bf p}_1 }( {\bf r},{\bf r}) \,
\exp - \frac{i}{\hbar}  ( {\bf Q \cdot r}) \,  d^3r 
\\ \nonumber
&=& A( {\bf p}_1, {\bf Q}-{\bf p}_1 )\,\tilde \phi_t ({\bf Q}-{\bf p}_1) ~ .
\end{eqnarray}
Here the integration over the coordinate with the 
weight $\exp - \frac{i}{\hbar}  ( {\bf Q \cdot r})$
ensures that the total momentum of the products
of the nuclear reaction is equal to ${\bf Q}$.

Let us now take the effect of TEC into account. We need
firstly to use the wave function (\ref{Psi}) which
describes the
amplitude of the event in which the inelastic collision
happens at the point ${\bf r}$ after TEC.
Secondly, we need to
multiply this amplitude by the 
amplitude of the nuclear reaction $A ( {\bf p}_1'', {\bf p}_2'' )$
and, additionally, 
by the factor $\exp-\frac{i}{\hbar}{\bf Q}\cdot{\bf r}$
(that again ensures that the products of the nuclear reaction have the 
total momentum ${\bf Q}$). Integrating the obtained result
over the coordinate, we derive the following expression for the 
matrix element,
\begin{equation}\label{3M}
\delta^{(TEC)} M ( {\bf Q} ) =  \int \, A ( {\bf p}_1'', {\bf p}_2'' ) \,
\delta^{(TEC)} \Psi ( { \bf r }, { \bf r } ) \, \exp - \frac{i}{\hbar}  
( {\bf Q \cdot r}) \, \,  d^3r ~.
\end{equation}
The procedure used to derive this result
is quite similar to  the case of 
the direct collision (\ref{M}). The only important distinction
is the wave function, which for the 
case (\ref{3M}) accounts for TEC.
The momenta 
${\bf p}_1'', {\bf p}_2''$ of the projectile and target nuclei
arise due to TEC and, henceforth, differ significantly
from the initial momenta ${\bf p}_1, {\bf p}_2$. This fact
drastically changes the exponential factor in (\ref{tun}), as is
discussed below.

\section{Probability of  nuclear reaction}
\label{proba}

To calculate the total probability of the nuclear reaction, one needs 
to fulfill integration of the square of the amplitude
over all possible momenta of the nuclear
reaction products in the final state.
For the direct collision this gives
\begin{eqnarray}\label{wdir}
&&W^{(DIR)} \propto  \int 
\left| M ^{(DIR)}( {\bf Q} ) \right|^2 \frac{d^3 Q}{(2\pi \hbar)^3}=
\\ \nonumber
&&\int \left| A ( {\bf p}_1, {\bf Q}-{\bf p}_1 ) \, 
\tilde \phi_t( {\bf Q}-{\bf p}_1 ) \right|^2 \frac{d^3 Q}{(2\pi \hbar)^3}=
\left| A ( {\bf p}_1, {\bf 0})\right|^2~.
\end{eqnarray}
Deriving the last identity we took into account the fact that
according to (\ref{p2<p1}) the Fourier component of the target wave function
is localized in a region of small momenta.
Eq.~(\ref{wdir}) indicates that the probability of 
the nuclear reaction initiated by the direct
collision in the environment is equal to the 
probability of the same event in the vacuum.

Consider now the probability of the nuclear reaction initiated by TEC
\begin{eqnarray}\label{d3Q}
W^{(TEC)} \propto
&& \int \left | \delta^{(TEC)} 
M ( {\bf Q} ) \right|^2 \frac{d^3 Q}{(2\pi \hbar)^3}=
\int \left|
A ( {\bf p}_1'', {\bf p}_2'' ) \,
\delta^{(TEC)} \Psi ( { \bf r }, { \bf r } ) \right|^2 \,d^3r =
\\ \nonumber
&&\int 
\exp \left( - \frac{ 2 \pi e^2\,Z_1 Z_2 }{\hbar v_{12}''}\right)
\left|
B ( {\bf p}_1, {\bf p}_2 )
\frac{f_{pt}(\theta_{pt})}{ R({\bf r}) }\,
\frac{  f_{pA}(\theta_{pA})}{ |{\bf r}-{\bf r}_A| }\, 
\frac{ f_{tB}(\theta_{tB})}
{ |{\bf r}- {\bf r}_B |}\, \phi({\bf r}_i) \right|^2 \,d^3r~.
\end{eqnarray}
Here the first identity takes into account (\ref{3M}) rewriting
the integration over the momenta $d^3Q$ 
as integration over the point of the final
inelastic collision $d^3r$. The second identity 
presents the wave function
$ \delta^{(TEC)}\Psi ( { \bf r }, { \bf r } )$ explicitly using
(\ref{Psi}), (\ref{sc}), and  (\ref{tun}).
The velocity in the exponential factor, 
\begin{equation}\label{v12}
v_{12}'' = |{\bf v}_1''-{\bf v}_2''| = 
\left|\frac{ {\bf p}_1'' }{m_p}-\frac{ {\bf p}_2'' }{m_t}\right|,
\end{equation}
depends on the momenta  ${\bf p}_1'',~{\bf p}_2''$
of the projectile and target nuclei after TEC.

Let us show that
the major contribution to the integral over $d^3 r$ in (\ref{d3Q})
originates from the vicinity of a segment 
which connects ${\bf r}_A$ and ${\bf r}_B$.
Consider instead of (\ref{d3Q}) an important relevant
integral
\begin{equation}\label{I}
I = \int \frac{ 1 }{ | {\bf r} - {\bf r}_A |^2\
| {\bf r} - {\bf r}_B |^2 } \,
\exp \left( -\frac{ 2 \pi e^2\, Z_1 Z_2}{\hbar v_{12}''} \right) 
\,d^3 r~.
\end{equation}
Let us choose cylindrical coordinates with their origin at ${\bf r}_A$ and the
axis pointing along the vector ${\bf r}_{BA}$. 
If we  call $z$ the distance from the origin in the
direction of this axis and $\vec \rho$ the radius vector 
in the orthogonal plane, then the radius vector
can be written as  ${\bf r} = (\vec \rho,z)$.
The projectile  and target nuclei after TEC
move along the segment ${\bf r} - {\bf r}_A$
and ${\bf r} - {\bf r}_B$, respectively.
Therefore the velocities in the cylindrical coordinate
can be written as
\begin{eqnarray}\label{v}
{\bf v}_1'' &=& 
v_1'' \frac{(\vec \rho,z)} {\sqrt{ \rho^2+z^2 } },
\\ \nonumber
{\bf v}_2'' &=& 
v_2'' \frac { (\vec \rho , z- r_{AB} ) }{ \sqrt{ z^2 + (r_{AB}-z)^2}}~. 
\end{eqnarray}
Expanding in powers of $\rho/r_{AB}$, one presents the collision velocity  
as a function of the coordinates,
\begin{equation}\label{v12''}
v_{12}'' = (v_1''+v_2'') \,
\left[ 1 - \frac{1}{2} \frac{v_1'v_2'}{(v_1'+v_2')^2} \,
\frac{r_{AB}^2\, \rho^2 }{z^2\,(r_{AB}-z)^2} \right|.
\end{equation}
We assume in (\ref{v}) that $z$ belongs to the segment 
$0 < z < r_{AB}$. In this 
region the collision velocity is large, therefore its contribution
is significant. Outside this region, when either $z<0$ or
$z>r_{AB}$, the collision velocity is small, drastically reducing
the integrand in (\ref{I}).

Substituting (\ref{v12''}) in (\ref{I}), one finds
\begin{equation}\label{Is}
I =  
\int_0^{r_{AB} }dz \int d^2 \rho 
\exp \left( - \frac{ 2 \pi e^2\,  Z_1 Z_2}{\hbar v_t }\right)\,
\frac{ 1 }{ | {\bf r} - {\bf r}_A |^2 | {\bf r} - {\bf r}_B |^2  } \,
\exp  
\left[- \pi \xi \, \frac{r_{AB}^2 \rho^2}{z^2\,(r_{AB}-z)^2}  \right]~,
\end{equation}
where $v_t$ is the sum of velocities,
\begin{equation}\label{vm}
v_t = v_1''+v_2'' =   v_1'+v_2'~.
\end{equation}
The last identity here is valid because we assume that
the collisions with
atoms A and B are elastic ones (or quasielastic ones, if we have in mind
that ionization of heavy atoms can occur).
The dimensionless parameter $\xi$, defined as
\begin{equation}\label{xi}
\xi =  \frac{e^2\, Z_1 Z_2 }{\hbar}\,\,
\frac{v_1'v_2'}{v_t^3}~, 
\end{equation}
governs  integration  over $d^2\rho$ in (\ref{Is}).
We consider low collision energies, therefore $\xi$ can be assumed to
be so large that  
\begin{equation}\label{gg}
\pi \xi \gg 1~.
\end{equation}
This inequality guarantees that separations from the
$z$ axis which are essential in the integrand in 
(\ref{Is}) are small compared to the separation between nuclei
$A$ and $B$,
\begin{equation}\label{rholl}
\rho \ll r_{AB}~.
\end{equation}
All essential parameters which govern the wave function
(including $v_1',v_2'$, {\em etc})
depend, generally speaking, on the final point. 
However, in the vicinity of the
$z$ axis their dependence on $\vec \rho$ becomes insignificant,
the only important dependence of the integrand on $\vec \rho$
is the factor $\rho^2$ in the exponent.
This  greatly simplifies 
integration over $d^2 \rho$ in (\ref{Is}), which results in
\begin{equation}\label{Isf}
I =  \int  \frac{1}{\xi r_{AB}}\, 
\exp \left( -\frac{ 2 \pi e^2\,Z_1 Z_2}{ \hbar v_t }\right)\,dz~.
\end{equation}
A similar method applied to (\ref{d3Q}) gives
\begin{equation}\label{d3Q'}
W^{(TEC)} \propto
\int 
\left| A ( {\bf p}_{1}'', {\bf p}_{2}'' )\right |^2 
\frac{ \left| f_{pt}({\theta_{pt}})\, f_{pA}( \theta_{pA} )
\,f_{pB}( \theta_{pB} )\right|^2 }
{R^2({\bf r}) r_{AB}^2}\, \left| \phi_t ( {\bf r}_i) \right|^2\,dz~. 
\end{equation}
Here the momenta are defined for the head-on collision 
which happens on the segment connecting ${\bf r}_A$ and ${\bf r}_B$,
\begin{eqnarray}
{\bf p}_{1}'' = -p_1' \frac{ {\bf r}_{AB} }{r_{AB}}~,~~~
{\bf p}_{2}'' = p_2' \frac{ {\bf r}_{AB} }{r_{AB}}~.
\end{eqnarray}
The scattering amplitudes in (\ref{d3Q'}) 
are also defined for the head-on collision.

Notice  that  the probability 
(\ref{d3Q'})  is proportional to the density 
$\left| \phi_t ( {\bf r}_i) \right|^2$
of the target nucleus at the point of initial
collision.  To make the probability bigger, we
need, therefore, to distinguish the case when the target
nucleus is well localized in the condensed matter or molecular
environment.
In this, most interesting for us, situation,
further simplification of (\ref{d3Q'})  becomes
possible.
When the target nucleus is localized
we should suppose that the relevant 
classical trajectories, which describe
propagation of the projectile and target
nuclei, start from some point in the vicinity of the localization point.
Our previous results indicate 
that the trajectories  finish at some point on the segment
$r_{AB}$. These facts altogether
fix the trajectories for the projectile and target nuclei uniquely.
As a result, the angles in the scattering amplitudes
as well as all velocities $v_1',v_2',v_t$ 
in the integrand in (\ref{d3Q'})
can be considered as 
parameters defined for the given classical trajectories.
In this case, 
the probability of the nuclear reaction
initiated by TEC $ W^{TEC}$, Eq.~(\ref{d3Q'}), can be presented as
\begin{equation}\label{W3}
W^{(TEC)} \propto
{\cal K} \left| A ( {\bf p}_{1}'', {\bf p}_{2}'' )\right |^2 ~,
\end{equation}
where the coefficient is
\begin{equation}\label{k}
{\cal K}=\frac{1}{\xi}
\frac{  \left| f_{pt}({\theta_{pt}})\, f_{pA}( \theta_{pA} )
\,f_{pB}( \theta_{pB} )\right|^2  }
{ R^2({\bf r}) r_{AB}^2 }
\, \int \left| \phi_t ( {\bf r}_i) \right|^2\,dz~.
\end{equation}
The factor ${\cal K}$, that appears in (\ref{W3}) 
and is defined in (\ref{k}), is the most important result 
of the calculations above.
It can be conveniently rewritten in terms of 
the differential cross sections which describe TEC,
\begin{equation}\label{kfin}
{\cal K}\Rightarrow {\cal K}_{AB}= \frac{1}{\xi}\,
\frac{ d \sigma_{pt}}{d \Omega}( \theta_{pt})\,
\frac{ d \sigma_{pA}}{d \Omega}( \theta_{pA})\,
\frac{ d \sigma_{tB}}{d \Omega}( \theta_{tB})\,
\frac{ 1 }{R^2({\bf r}) r_{AB}^2 }
\int n_t({\bf r}_i)\,dz~.
\end{equation}
We introduce here
the density of the target nucleus $n_t({\bf r}_i)= |\phi_t ({\bf r}_i)|^2$
at the point of the initial elastic collision ${\bf r}_i$.
In order to fulfill integration in (\ref{kfin})
over the $z$ coordinate of the point of the final collision,
one needs to remember that 
${\bf r}_i$ is a function of
the point of the final collision ${\bf r}$,
see discussion in Section \ref{wf}
and Appendix A.  This integration 
cannot be simplified any further in the general case,
but it presents no difficulties for numerical estimations, 
see example in Section \ref{num}.
The consideration above ignored the temperature of the target.
If we assume that the temperature is lower than the projectile
energy, $T \ll E_1$, then by slightly modifying
the approach above, we arrive at the same result
(\ref{kfin}), with the density $n_t({\bf r}_i)$ a function of
the temperature.

Eq.~(\ref{kfin}) shows that each scattering process 
gives a factor into the probability which
is identical to the elastic cross section; that 
result could have been anticipated without any calculations.
Having established this, one could  believe that
an estimate for the pre-exponential factor should look like
${\cal K} \sim 
\frac{ d \sigma_{pt}}{d \Omega}(\theta_{pt})\,
\frac{ d \sigma_{pA}}{d \Omega}(\theta_{pA})\,
\frac{ d \sigma_{tB}}{d \Omega}(\theta_{tB}) 1/r_{AB}^6 ~,$
because the atomic separation $r_{AB}$  seems  to be the only relevant 
parameter in the problem. However, our calculations
(\ref{kfin}) reveal that there exists an additional
enhancing factor $k$,
\begin{eqnarray}\label{en}
{\cal K} &\simeq&  \frac{k}{\xi r_{AB}^6}\,
\frac{ d \sigma_{pt}}{d \Omega}(\theta_{pt})\,
\frac{ d \sigma_{pA}}{d \Omega}(\theta_{pA})\,
\frac{ d \sigma_{tB}}{d \Omega}(\theta_{tB}) ~, \\ \label{kk}
k &=& r_{AB}^2 \int n_t({\bf r}_i)dz \sim r_{AB}^2 n_t^{2/3}(0) \simeq
\frac{r_{AB}^2}{a^2}\simeq 100~,
\end{eqnarray}
where $n_t(0) = 1/a^3$ is the maximum density of the target nucleus 
at the point of localization, and $a$ is the corresponding
length which describes the region of localization.  The latter one 
depends on the
amplitude of vibrations exhibited by the target in the
considered  condensed matter or molecular environment.
This length can usually 
be estimated as $\sim 10\% $ of an atomic separation
\cite{ziman}, which justifies the last identity in 
(\ref{kk}). 
\footnote{
Note that in the derivation of (\ref{kfin}),
the length of localization $a$ was assumed to be sufficiently large, 
$a^2 \gg \frac{d \sigma_{pA}}{d \Omega},~\frac{d \sigma_{tB}}{d \Omega}$,
which is true for sufficiently high energy.
}
Eqs. (\ref{en}), (\ref{kk}) predict that
\begin{equation}\label{4/3}
{\cal K} \propto r_{AB}^{-4} \propto n^{4/3}~,
\end{equation}
where $n$ is the density of the condensed matter environment.
We will use this fact for estimations described below in Section \ref{con}.

We consider above a chain of  scattering processes with the participation 
of atoms A and B, indicating this fact in the notation
${\cal K}_{AB}$ in (\ref{kfin}).
If the process happens 
in large molecules, or in condensed matter, 
there may be several heavy atoms located close to the
target nuclei.  Each of them provides the possibility for the
necessary elastic rescattering. 
Therefore, if we want
to consider all possible rescattering processes we need simply to
add the contributions arising from scattering on different neighbor nuclei.
If the nucleus $A$ participates in the process
then it gains a large momentum. Therefore, the
contributions from scattering on different nuclei do not interfere, and we
need to sum the probabilities.
The result again can be presented with the help
of (\ref{W3}), where the coefficient is
\begin{equation}\label{sumk}
{\cal K} = \sum _{A,B}{\cal K}_{AB}~.
\end{equation}
The summation here runs over all neighbor nuclei of the environment.

Consider the ratio of  the probability of the
nuclear reaction initiated
by TEC to the probability of the direct event, 
\begin{equation}\label{F}
{\cal R} = \frac{ W^{(TEC)} } { W^{(DIR)} }~.
\end{equation}
Calculating this ratio 
with the help of (\ref{wdir}), (\ref{W3}),
one should remember that the probability of the 
nuclear reaction
includes the sophisticated matrix elements 
which describe the proper nuclear
processes and are defined as the amplitude
$B ( {\bf p}_1, {\bf p}_2 )$, Eq.~(\ref{tun}).
These latter factors vary on the scale of nuclear energies
which are much higher than the collision energy. 
Therefore,
with high accuracy 
these nuclear matrix elements can be considered as energy-independent
constants, which cancel out in the ratio (\ref{F}). 
Having this fact in mind,
we find the following 
expression for the ratio ${\cal R}$,
\begin{equation}\label{FF}
{\cal R}  = {\cal K}\, \exp \left[ \frac{2 \pi e^2\,Z_1 Z_2}{\hbar} 
\left(\frac{1}{v_1} - \frac{1}{v_t}\right)\right]~.
\end{equation} 
Here  the factor ${\cal K}$ is defined in (\ref{kfin}), (\ref{sumk}).
Note an important inequality $v_m > v_1  $, which guarantees
that the ratio ${\cal R}$ {\em increases exponentially}
for low energies $E_1$.

The applicability of the developed theory is restricted 
by several bounds. 
One of them comes from within the theory.
Inequality (\ref{gg}) requires that
  $\xi$ is large, thus putting a bound for the energy
from above.
For an important example of the collision of
two deuterons, this inequality gives $v_1 < (e^2/\hbar)\, \pi/(4 \sqrt 2) $, 
resulting in $E_1 \le 15$ keV. 
From below, the energy in our approach is obviously 
restricted by 
a typical potential energy in a crystal or molecular
environment, that necessitates (\ref{qlarge}).

\section{Screening} \label{screen}
Screening  plays an essential role
in low energy collisions of nuclei in condensed matter  \cite{Ichimaru}.
Let us summaries briefly relevant facts.
When the two  colliding nuclei come sufficiently close, their
influence on the electrons of the environment
can be described by their total
Coulomb charge $Z= Z_A+Z_B$.
This charge for low energy collision
creates discrete atomic-like energy levels for electrons. 
The electrons of the environment can 
populate these levels and
form an effective atom  around the {\em combined} nucleus.
This is the strongest possible  
response which the electrons can exhibit in screening the 
Coulomb field.
An occupation of discrete levels in the combined atom
results in energy production. 
The maximal energy produced in this process is equal to
\begin{equation}\label{energy}
U = E_B(A_Z) - Z \cdot \Delta~,
\end{equation}
where  $E_B(A_Z)$ is the total binding energy of the atom
$A_Z$ that has $Z$ electrons surrounding
the combined nucleus with nuclear charge $Z$, and
$\Delta$ is the binding energy for each electron
in the given condensed matter environment.
The energy variation $U$ 
can be transferred into the 
kinetic energy of the colliding nuclei.
This transfer may be expected to be efficient
for small collision velocity. 
Eq.(\ref{energy}) shows that  screening
makes the collision energy higher $E_{col} \rightarrow
E_{col} + U$. This, in turn, makes
the relative velocity larger
\begin{equation}\label{rel}
v_{12} \rightarrow v_{S} = \sqrt{ v_{12}^2 + 2 U/\mu}~,
\end{equation}
thus increasing
the probability to penetrate through the Coulomb barrier
(\ref{prob}) and  boosting the nuclear event.
Screening in this description is described by the
screening potential $U$. The larger $U$, the stronger 
is the screening  and the higher is the
probability of the nuclear reaction. 
It is clear that the given consideration 
may overestimate a role of screening 
because, firstly, a probability to populate
the ground state of the effective atom  
should, generally speaking, be less than unity, and, secondly,
because the energy production should not necessarily
be transferred solemnly into the kinetic energy of the colliding nuclei,
but may be used for excitation of 
the environment as well. 
These two factors could make screening potential
less than what is predicted by (\ref{energy}).
Accurate account of all relevant factors is difficult and,
probably, due to this reason Ref. \cite{Pd}
uses $ U $ as a parameter 
to fit the experimental data.

The above discussion shows that
screening results in an increase of the relative velocity.
This effect takes place in the direct
collision as well as in the TEC process.
To take screening into account we need
to increase velocities in both these collisions.
Applying this procedure to the relative probability 
(\ref{FF}) of the nuclear process initiated
by TEC we find 
\begin{equation}\label{Rf}
{\cal R} \rightarrow
{\cal R}_S  = {\cal K}\, \exp \left[ \frac{2 \pi e^2\,Z_1 Z_2}{\hbar} 
\left(\frac{1}{v_{S}} - \frac{1}{v_{t, \,S}}\right)\right]~,
\end{equation} 
where $v_{1,\,S}$  and  $v_{t, \,S}$ are defined in accord with
(\ref{rel})
\begin{eqnarray}\label{v1S}
v_{1,\, S} &=& \sqrt{ v_1^2 + 2 U/\mu}~, \\ \label{vtS}
v_{t, \,S} &=& \sqrt{ v_t^2 + 2 U/\mu}~.
\end{eqnarray}
It is clear that screening produces stronger impact
for the direct collision  because
the relative velocity
in this case is smaller 
than in the collision initialed by TEC.
Therefore screening should  reduce
the relative probability  ${\cal R}_S$
(\ref{Rf}) in comparison with its non-screened value
${\cal R}$
(\ref{FF}), see Section \ref{num} for more details.

\section{Applications for deuterium-deuterium fusion}
\label{num} 
Let us apply the results obtained above to the
important case of deuterium fusion 
${\rm d}+{\rm d} \rightarrow {^3{\rm H}}+ {\rm p}$
in a collision of the
projectile deuterons  with target deuterons implanted in a solid.
Eq.~(\ref{kfin}) shows that we need to estimate the
cross sections for TEC. 
For the initial deuteron-deuteron elastic collision one can use the
conventional Rutherford formulae
\begin{equation}\label{Ru}
\frac{d \sigma_{pt}}{d \Omega}(\theta_{pt}) = 
\left( \frac{e^2}{2 E_1}\right)^2 
\frac{1}{ \sin ^4 (\theta_{pt}/2)}~,
\end{equation}
where $E_1$ is the energy of the projectile deuteron and 
$\theta_{pt}$ is the scattering angle 
in the center-of-mass reference frame.
It should be noted that the target 
deuterium in a condensed matter environment
can form a neutral atom;
the originally charged projectile nuclei 
propagating through the environment
can also pick up an electron becoming  neutral as well.
Thus, during the deuteron collision the charges of the nuclei
are partly screened by the deuterons' electrons.
This fact, however,  does not change 
estimate (\ref{Ru}) which 
assumes the pure Coulomb interaction between the 
nuclei. The necessary significant variation of velocities
of the deuterons
requires that their collision should be sufficiently hard, 
i.e. happen 
with  small impact parameter $b$, $b \simeq e^2/E_p$.
Therefore, for energies higher than
the atomic energy, $E_p \gg 27.21 $ eV, the impact parameter
must be much smaller than the Bohr radius, $b \ll a_0$.
The smallness of the impact parameter shows that
screening produces a small effect for deuterium scattering.

The same argument can be applied to the
collision of a deuteron with a heavy atom.
Electrons located at the deuteron nucleus or 
in an outer shell of the heavy atom cannot 
produce any significant effect 
because major events
during the scattering happen for small separation between the
deuteron nucleus and the nucleus of the heavy atom.
\footnote{This argument is in line with
the discussion of the role of ionization
during collisions, see Section \ref{wf}.
The collision with the heavy atom can, generally speaking, 
be accompanied by ionization of the heavy atom or the  
deuteron atom.
However, the ionization process 
involves an outer shell of the heavy atom or the deuteron atom.
This means that the ionization
happens when the deuteron nucleus
is separated by a large distance $\sim a_0$ from the heavy atom nucleus.
In contrast, large variation of the velocity of the deuteron nucleus
happens at small separations, where ionization does not play a role.
We conclude again that scattering with ionization can be considered as 
a quasi-elastic process which does not affect significantly the 
collision with large variation of the
deuteron velocity.}

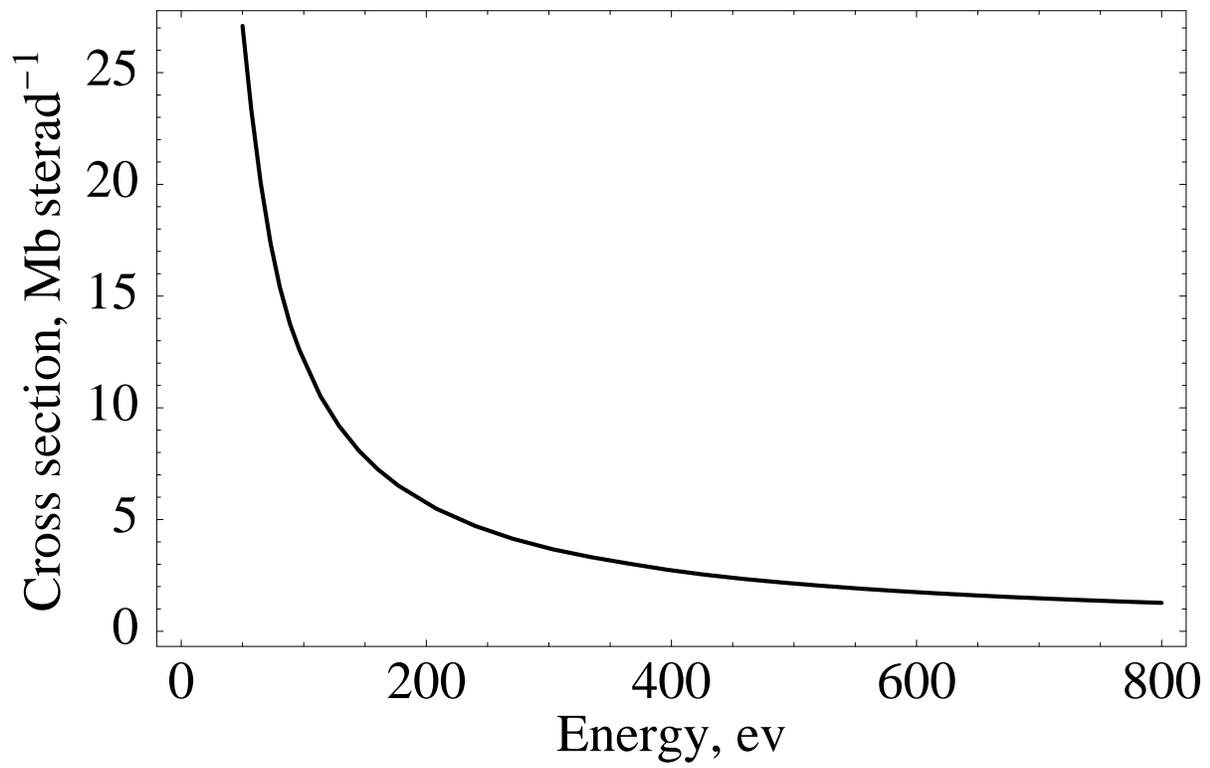
\begin{figure}
\caption{ \label{cross}
The differential cross section 
$\frac{d \sigma}{d \Omega}(\theta)$ for  elastic scattering
of deuterons by gold atoms calculated with the help of Eq.~(\ref{classigma})
for the scattering angle $\theta =135^0$.}
\end{figure}

The above arguments are valid for an outer shell of the heavy atom.
In contrast, electrons of inner shells are localized  closer to the nucleus
and therefore produce a strong effect,
providing screening for the Coulomb field created by the heavy nuclei
for those small separations which are responsible for the scattering.
The screening can be conveniently taken into account
in the semiclassical approximation, the validity of which
is justified by the large mass and energy of a deuteron.
In this approach one needs firstly to calculate
the deuteron scattering angle as a function of the
impact parameter $\rho$ \cite{Landau1},
\begin{equation}\label{impact}
\theta = \left| \pi - 2 \int_{r_{\rm min} }^\infty
\frac{ \rho }
{ r^2 }\,
\left( 1 - \frac{\rho^2}{r^2} - \frac{ 2 U(r) }{ m v_1'^2} \right)^{-1/2} dr
\right|~.
\end{equation}
Then the cross section is found from the conventional
relation \cite{Landau1}
\begin{equation}\label{classigma}
\frac{d \sigma}{d\Omega}(\theta) = 
\frac{\rho}{\sin \theta}\frac{d \rho}{d \theta}~.
\end{equation}
We perform these calculations using 
the well-known Thomas-Fermi atomic potential \cite{Landau}
for the atomic potential energy $U(r) = U_{TF}(r)$. 
Fig.~\ref{cross} presents results for
deuteron scattering from gold atoms.
Our calculations for other atoms give similar results.
The cross section is comparable
with an area with the atomic dimension $a_0^2$ for low energies and
monotonically decreases with energy.

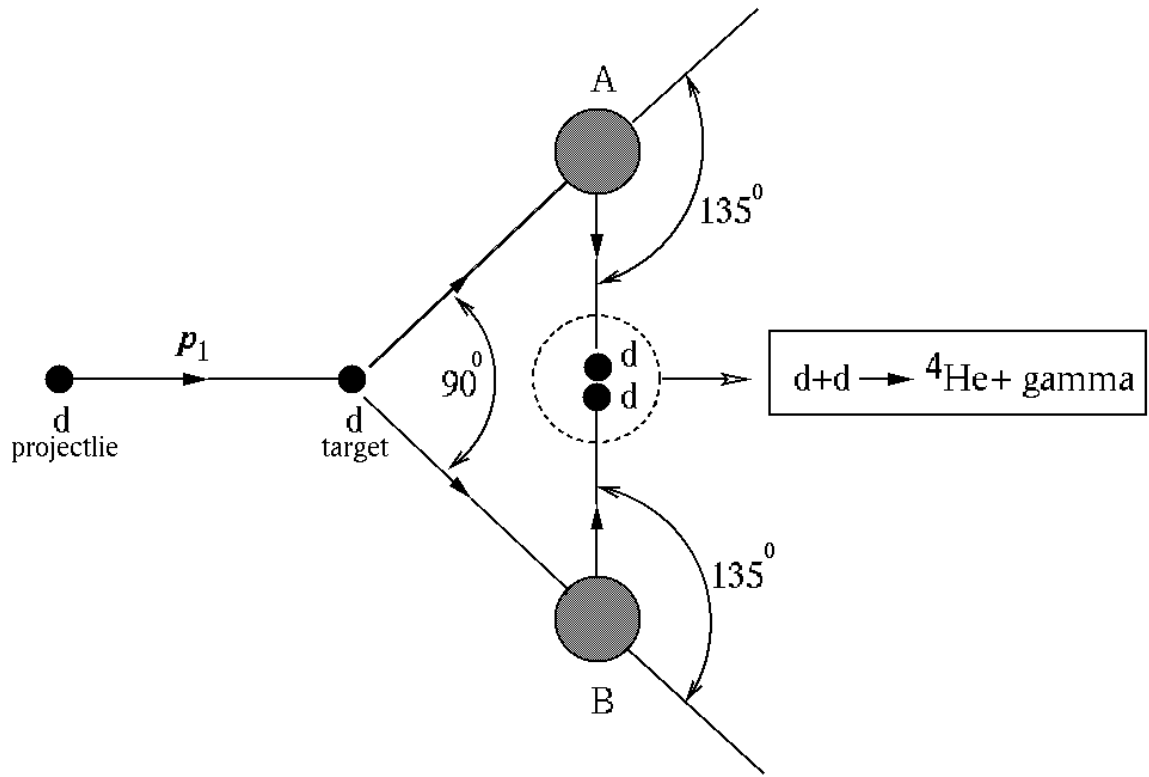
\begin{figure}
\caption{ \label{dd}
Three elastic collisions for deuteron
fusion. 
The initial momentum  of the
projectile deuteron ${\bf p}_1$ is orthogonal
to the segment $R_{AB}$. The target deuteron
is located so that the final inelastic collision
happens with  velocity
$v_{12} = \sqrt2\, v_1$.
This is the largest possible collision velocity, therefore
the considered geometry is the most suitable one.}
\end{figure}

To proceed further we need to fix the geometry of the collision.
Let us assume that the momentum of the projectile deuteron ${\bf p}_1$
is orthogonal to the segment $AB$, i.e. ${\bf p}_1 \cdot {\bf r}_{AB}=0$.
Assume also that the target deuteron is located in the vicinity
of the point most convenient for this case, 
${\bf r}_0 = {\bf r}_{AB}/2 - 
({\bf p}_1/p_1) r_{AB}/2$. Then the scattering angles are
\begin{eqnarray}\label{pt}
\theta_{pt} = 90^o\, ,~~~ \theta_{pA} = \theta_{tB} = 135^o ~.
\end{eqnarray}
In this geometry, shown in Fig.~\ref{dd}, 
the final inelastic collision happens at the point
${\bf r}_{AB}/2$.
To be specific, we assume that the atoms A and B are the atoms of gold
(though this assumption is not essential, other heavy
atoms would produce a similar effect).

The velocities after the first deuteron-deuteron collision
are $v_1'=v_2'= v_1/\sqrt{2}$, which makes the collision velocity for 
the final inelastic collision (\ref{vm}) 
\begin{equation}\label{fcv}
v_t = \sqrt{2} \,v_1~.
\end{equation}
From this we deduce for the ratio (\ref{Rf})
\begin{eqnarray}\label{RR}
{\cal R} = {\cal K} \exp \left[ \frac{2\pi e^2}{\hbar} \,
\left( \frac{1}{ v_{1,\,S} } - \frac{1}{v_{t,S} }\right) \right]~,
\end{eqnarray}
where
\begin{eqnarray}\label{KK}
{\cal K} =  C\, \frac{1}{\xi} \frac{k}{r_{AB}^6} \, 
\frac{e^4}{E_1^2}
\left[ 
\,  \frac{ d\sigma_{pA}}{d\Omega}(135^o) \right]^2 
\end{eqnarray}
and $C = 0.7$ is a numerical factor. 
\footnote{
It is 
calculated as a product $C=2\cdot(2/3) \cdot c$
in which the coefficient 2
takes into account the fact that the sum in (\ref{sumk}) 
includes two
possible trajectories for the pair of nuclei 
which lead each one of them towards either atom A or B,
the coefficient 2/3 originates from  Eq. (\ref{t=t}),  
and the numerical factor $c = (\sqrt2+1/2)^{-1}=0.522$ arises 
from the integration over $ dz = c dz_i $ in Eq. (\ref{kfin})}.
For the considered case one finds that the parameter $\xi$
defined by (\ref{xi}) is equal to 
\begin{equation}\label{xidd}
\xi = \frac{1}{4\sqrt2}\, \frac{ e^2 } {  \hbar v_1 }~.
\end{equation}
The velocities
$v_{1,\,S}$ and $v_{t,\,S}$ in the exponent of (\ref{RR})
are defined in (\ref{v1S}),(\ref{vtS}), thus taking screening into account.

Let us estimate $r_{AB}$ as a typical atomic separation, say the
separation between atoms of gold metal, 
$r_{AB}=407.82\,{\rm nm} = 7.71$ au. 
The enhancing coefficient $k$ according to
(\ref{kk}) is $ k \simeq 100 $.

Fig. \ref{Rfactor} shows the results of our calculations 
for the ratio ${\cal R}$ (\ref{RR}) of the
probability of the nuclear reaction
initiated by TEC to the probability of the direct collision.
The coefficient ${\cal K}$ was
calculated from (\ref{KK}) as explained above.
The screening potential $U$, which influences the velocities 
$v_{1,\,S}$ and $v_{t,\,S}$ in 
(\ref{v1S}),(\ref{vtS}) was found experimentally in Ref.\cite{Pd}
for the case of Au
to be $U = 22.8 \pm 11.0 $ eV. 
Estimation (\ref{rel}) predicts a higher value. 
Note that the combined atom in the case considered 
is He, its binding energy is E(He($1s^2$)) = 79.0 eV. If we assume that
$\Delta \simeq 13.6$ eV (binding energy of the Hydrogen atom), 
than  (\ref{rel}) gives
$U = 51.8 $ eV. Deviation of this result from
the experimental data prompts us 
to present
in  Fig. \ref{Rfactor} data which show
variation of the enhancing factor ${\cal R}$
both on the collision energy and
the screening potential  $U$.
We see that for collision energies below $\sim 500$ eV
the  ratio ${\cal R}$ becomes large, 
 ${\cal R}\ge 1$, showing that the TEC 
makes the probability of the nuclear reaction
bigger than in the direct inelastic collision. 
For lower energies this enhancement becomes stronger.
Screening makes the nuclear reaction more probable, but
it results in the decrease of the relative probability 
probability ${\cal R}$ for processes initiated by TEC. 
This damping effect arises because screening
manifests itself  stronger for the direct process,
in which the collision energy is lower.
The stronger screening, 
i.e. the higher the potential $U$ is, 
the less important role plays the TEC mechanism.

The analysis above assumes that the solid state environment
keeps the target deuteron localized in the center of
a square which is cornered by heavy atoms, and
the beam of the projectile deuterons moves parallel
to some side of this square. This geometry is the most
favorable because it provides 
an enhancement factor (\ref{en}). For an arbitrary
geometry, the enhancement (\ref{en}) does not take place,
making the probability of TEC lower by
a factor of $\sim 100$.

We conclude that the fusion below 500 eV 
is strongly influenced by the TEC mechanism.

\section{Conclusion}
\label{con}

We have demonstrated that there is a boost in the probability of a 
nuclear reaction, in a condensed matter or molecular environment, 
when three elastic collisions (TEC) precede the inelastic collision 
of two nuclei. 
When operative, the
TEC mechanism provides an {\em exponential enhancement}
for nuclear reactions with low collision-energy.
One should remember, however, that this boost
does not fully compensate for the suppression
due to the Coulomb repulsion.
It accounts, roughly speaking, for only $30 \%$
($ 0.3 \approx 1-1/\sqrt 2 $, see (\ref{RR}) )
of the exponent which describes tunneling
through the Coulomb barrier, leaving $70 \%$
of this exponent intact.
Moreover, the compensation originates from a 
chain of several  (three) collisions which produces 
an additional pre-exponential, but nevertheless important, damping factor.
Additionally, screening also reduces the role played by TEC.
As a result of this we find that 
the probability for a nuclear reaction
is enhanced by TEC only
for sufficiently low collision energies.
An example of DD fusion discussed in Section \ref{num}
demonstrates that the effect of TEC manifests itself
strongly for energies 
below 500 eV, being negligible for higher energies. 
Obviously, in this low energy region
the probability of fusion is very low, 
by a factor of $10^{14}$ lower than for the energy 2 keV.
This should make direct measurement of the discussed effect
very difficult in the near future.
\footnote{
Hence, the discussed effect can not be directly responsible
for some unexplained anomalies in fusion 
which have been observed experimentally.}

Nevertheless, the effect  results
in a strong boost for low-energy nuclear reactions which may be
important for different applications.
One of them
can be looked for in laser-induced fusion, in which
ions can be accelerated by the laser field, see e.g. \cite{Hora}.
This acceleration can, generally speaking, 
create a flux of deuterons which move
from a surface inside a target exposed to a laser field.
If these ions have energies 
below $0.5$ keV, than
the TEC mechanism can boost the fusion.
One can deduce a practical conclusion from this.
To allow the TEC mechanism to manifest itself  
in laser-induced fusion,
the deuterium target should 
contain a sufficient amount of heavy atoms.

One can look for other applications of the TEC mechanism, 
for example, in astrophysics.

\appendix

\section{Wave function in the semiclassical approximation}

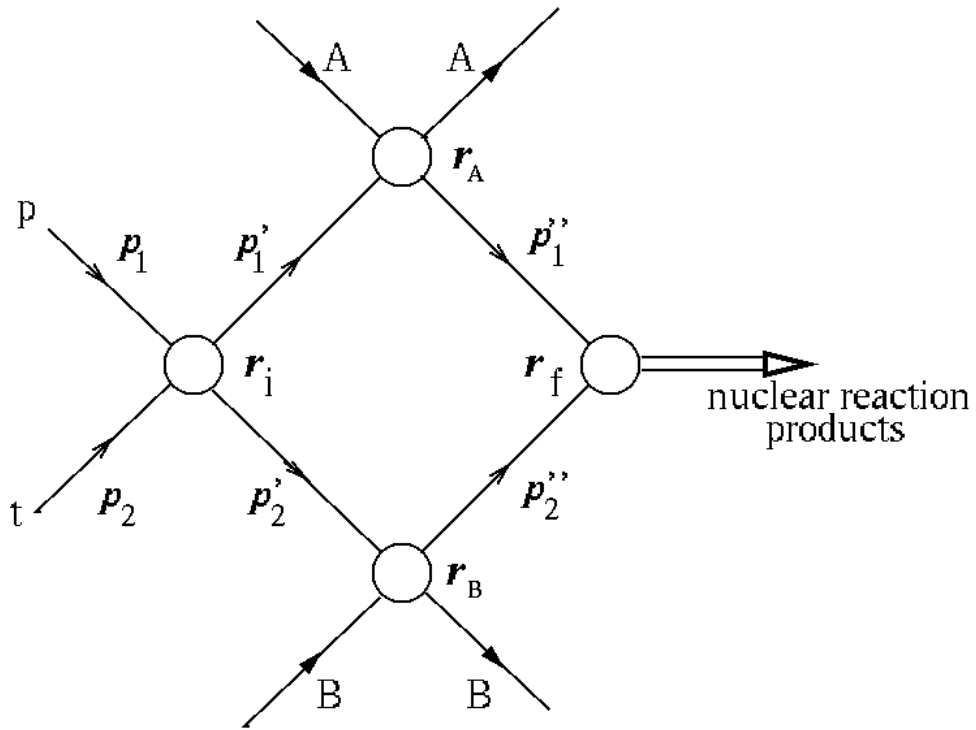
\begin{figure}
\caption{ \label{feynman_d}
The Feynman diagram which describes the mechanism of
three elastic collisions.
Solid lines with arrows in the intermediate states
show propagators of the
projectile and target nuclei.
They are marked by the corresponding momenta.
Open circles describe 
the three elastic scattering processes.
Each  scattering happens at some well defined
point: the initial one takes place at ${\bf r}_i$.
The scattering of the nuclei by atoms A and B 
occurs at the points  ${\bf r}_A$ and ${\bf r}_B$ 
where these atoms are located.
The final
inelastic collision takes place at ${\bf r}_f \equiv {\bf r}$. 
Since all these points  are well separated in space,
these collisions may be described by the 
physical scattering amplitudes (on mass-shell amplitudes).
Calculation of this Feynman diagram is presented in 
Appendix A.}
\end{figure}

Let us derive the wave function 
which describes the projectile and target nuclei
with TEC taken into account.
This goal can be achieved with the help of the Feynman 
diagram presented in Fig.~\ref{feynman_d}.
The solid lines with open arrows
on the diagram show  the projectile and 
target nuclei. Their propagation in the intermediate states
should be described with the help of the
corresponding Green functions $G_p,\,G_t$.
Since we assume that the energies of the nuclei are 
sufficiently high, see  (\ref{qlarge}),
we can use the free-particle
Green functions $G_a = G^{(0)}_a$, $a=p,t$.
In the coordinate-time picture the propagator
of a free particle is
\begin{equation} \label{Gi}
G_a^{(0)}({\bf r},t; {\bf r}',t')=
-\frac{i}{\hbar} \left( \frac{m_a}{ 2\pi i \hbar (t'-t) } \right)^{3/2}
\exp \frac{i}{\hbar} \left( \frac{m_a |{\bf r}'-{\bf r}|^2}{t'-t}\right)~.
\end{equation}
Open circles on the diagram describe interactions
which are responsible for TEC.  Let us use notation
$M_{ab}$ (with appropriate labels $ab= pt,\, pA,\,tB$)
to describe the corresponding matrix elements.
The important point is that each step of TEC happens inside some
localized region, each collision being well separated 
from other ones. 
As a result, each step of TEC can be described
by a physical (``on mass shell'') scattering amplitude.
The matrix elements $M_{ab}$ are related to the scattering amplitudes
$f_{ab}$ by conventional formulae,
\begin{equation}\label{fM}
f_{ab} = -\frac{m_{ab}}{2\pi\hbar^2} \, M_{ab}~,
\end{equation}
where $m_{ab}$ is a reduced mass of the pair of particles $ab$.
The scattering angles in these amplitudes 
are specified below.

The considered diagram Fig.~\ref{feynman_d} gives the following 
contribution to the wave function, 
\begin{eqnarray} \label{d3rdt} 
&& \delta^{(TEC)}\Psi({\bf r}_1,{\bf r}_2) = 
\int_0^{t_f}dt_A \int_0^{t_f}dt_B \int_{\max {(t_A,t_B) }} ^\infty dt_f
\int d^3 r_i \exp\frac{i}{\hbar}
({\bf P} \cdot {\bf r}_i + E t_f)
\\ \nonumber 
&\times &M_{pt} \,G^{(0)}_p({\bf r}_i,0; {\bf r}_A,t_A) M_{pA}\,
 G_p^{(0)}({\bf r}_A,t_A; {\bf r}_1,t_f)
                G_t^{(0)}({\bf r}_i,0; {\bf r}_B,t_B) \,M_{tB} \,
                G_t^{(0)}({\bf r}_B,t_B; {\bf r}_2,t_f)  \\ \nonumber
&\equiv&
\int_0^{t_f}dt_A  \int_0^{t_f}dt_B \int_{\max {(t_A,t_B) }} ^\infty dt_f
\int d^3 r_i  F \exp \frac{i}{\hbar} S~.
\end{eqnarray}
The integration here includes integration over the point of initial
collision ${\bf r}_i$ and integration over the
time variables. The latter describe the
times of the elastic collisions:
the projectile collision with the atom A ($t_a$),
the time of the target collision with atom B ($t_B$),
and the final time ($t_f$).
The quantities ${\bf P}={\bf  p}_1+{\bf p}_2$
and $E=E_1+E_2$ are the total momentum 
of the pair in the initial state and the corresponding energy.

We use in (\ref{d3rdt}) the ``time-coordinate''
presentation for the propagators which is convenient
for our purposes since locations of atoms A and B are fixed. 
Another, probably more widely used, presentation
provides the ``energy-momentum'' picture, 
in which propagations of  particles are described by
conventional energy denominators. We verified that
(\ref{d3rdt}) can be directly
derived from the energy-momentum presentation.

The quantity 
 $S$ in the exponent in (\ref{d3rdt})
originates from the phases of the four Green functions
(\ref{Gi}) combined with the phase factor 
${\bf P}\cdot {\bf r}_i +E t_f$,
\begin{eqnarray}\label{S}
S &=& S( {\bf r}_1,{\bf r}_2; t_A,t_B,t_f, {\bf r}_i)=
{\bf P}\cdot {\bf r}_i + E t_f  \\ \nonumber
&+& \frac{ m_p  | { \bf r }_A - { \bf r }_i |^2 }{ 2( t_A-t_i)^2 }+
\frac{m_p  | { \bf r } - { \bf r }_A |^2 }{2( t_{f}-t_A)^2}+
\frac{m_t  | {\bf r}_B - {\bf r}_i |^2 }{2( t_{B}-t_i)^2}+
\frac{m_t  | {\bf r} - {\bf r}_B |^2 }{2( t_{f}-t_B)^2}~.
\end{eqnarray}
One recognizes in this expression the classical action
which describes the free
motion of the two nuclei between the collisions.
The factor $F$ in (\ref{d3rdt}) describes all relevant
pre-exponential factors.

The heavy, high energy
nuclei have a wavelength much less than the atomic separation.
Therefore, propagation of the nuclei
in a condensed matter environment can be described semiclassically.
This means that  the integral in (\ref{d3rdt}) can be evaluated
using the saddle-point approximation. The saddle-point
position should be found from the conditions
\begin{equation}\label{gradS}
\frac{\partial S}{\partial {\bf r_i}}= 
\frac{\partial S}{\partial t_A}= 
\frac{\partial S}{\partial t_B}= 
\frac{\partial S}{\partial t_f}=0~,
\end{equation}
which coincide with the classical equations of motion for the two nuclei.
The classical equations specify the locations and moments
of time of TEC which are the functions
of the coordinates ${\bf r}_1,{\bf r}_2$. 
They specify also the momenta
of the projectile and target nuclei in the intermediate state.
These momenta allow one to find  the scattering angles that
govern the matrix elements $M_{ab}$ in (\ref{d3rdt}).

In the semiclassical approximation,
the final pre-exponential factor  is related to the action,
since
\begin{eqnarray}\label{intd6}
\int F \exp \frac{i}{\hbar} S \,\, dt_A dt_B dt_f d^3 r_i  =
\left( \frac{2\pi i}{\hbar}\right)^{3/2}\frac{F}{\sqrt{ \det S''}}~.
\end{eqnarray}
Here $S''$ is  the $6\times 6$ matrix of
second derivatives of the classical action over the six 
integration variables 
$t_A,t_B,t_f, {\bf r}_i$.
To simplify the formulae,
we restrict the following 
consideration to the specific
case when ${\bf r}_1={\bf r}_2 \equiv {\bf r}$.
In this case one  finds that the   matrix  $S''$ is 
\begin{eqnarray}\label{S''}
S'' = \left( 
\begin{array}{cccc}
2 E_1'\left(\frac{1}{t_A}+\frac{1}{ t_{fA} }\right)&
0&
-\frac{ 2E_1' }{t_{fA}}& 
\frac{ p_{p\beta}' }{t_A} \\
0&
2 E_2'\left( \frac{1}{t_B}+\frac{1}{ t_{fB} } \right)&
-\frac{ 2 E_2' }{ t_{fA} }& 
\frac{p_{t\beta}'}{t_B} \\
-\frac{ 2 E_1' }{t_{fA}}& 
-\frac{ 2 E_2'}{t_{fB}}& 
2\left( \frac{ E_1' }{ t_{fA} }+\frac{ E_2'}{ t_{fB} } \right)&
0 \\
\frac{ p_{p\alpha}' }{t_A}&
\frac{ p_{t\alpha}' }{t_B}&
0&
\delta_{\alpha\beta} \left( \frac{m_p}{t_A}+\frac{m_t}{t_B}\right)
\end{array}
\right)~,
\end{eqnarray}
where $\alpha,\beta =1,2,3$ describe the Cartesian projections of the vectors.
Calculating the determinant of the matrix (\ref{S''}) and 
substituting the result in  (\ref{intd6}) and
(\ref{d3rdt}),
we find the wave function which can be written in the following form
\begin{eqnarray}\label{psi3}
\delta^{TEC}\Psi({\bf r},{\bf r}) =
\frac{f_{pt}(\theta_{pt})}{ R({\bf r}) }\,
\frac{  f_{pA}(\theta_{pA})}{ |{\bf r}-{\bf r}_A| }\, 
\frac{ f_{tB}(\theta_{tB})}
{ |{\bf r}- {\bf r}_B |}\, \exp \frac{i}{\hbar} S 
\end{eqnarray}
The action on the classical trajectory can be presented
in a convenient, more familiar form
\begin{equation}\label{phasa}
S = {\bf P}\cdot {\bf r}_i + 
p_1'(|{\bf r} -  {\bf r}_A| +|{\bf r}_A - {\bf r}_i|) + 
p_2'(|{\bf r} -  {\bf r}_B| +|{\bf r}_B - {\bf r}_i|)~.
\end{equation}
The factor $R({\bf r}_i)$ in (\ref{psi3}) comprises 
the result of the calculation for $\det S''$,
\begin{eqnarray}\label{RS''}
R({\bf r}) = 
\left( v_{12}'\,^2 +\frac { m_p v_1'\,^2 t_{fB}+ m_t v_2'\,^2 t_{fA} }
{m_p t_{Bi} + m_t t_{Ai} } \sin^2 \gamma \right)^{1/2}
\frac{ m_p t_{Bi} + m_t t_{Ai} }{m_p+m_t}~,
\end{eqnarray}
where the angle $\gamma$ is defined  by
\begin{equation}\label{cos}
\cos \gamma = \frac{ {\bf v}_1' \cdot {\bf v}_2' }{ v_1'v_2' }~.
\end{equation}
The notation above assumes the usual convention 
$t_{ab}= t_a -t_b$.
Eq.~(\ref{RS''}) looks complicated, but it greatly simplifies
for a particular important case,
 $m_p=m_t,~t_{Ai}=t_{Bi}$, and $\cos \gamma =0$,
when it becomes
\begin{equation}\label{t=t}
R({\bf r}) = \sqrt{ \frac{3}{2} }\,r_{AB}.
\end{equation}
Otherwise, in the general case, one can make  an estimate 
$R({\bf r}) \sim r_{AB} $, which suffices for our purposes.

The wave function (\ref{psi3}), that takes into account TEC, 
is the final result of this section. To avoid confusion, let us 
note that it does not account for the Coulomb repulsion
of the nuclei when they finally collide inelastically.
This effect is taken care of separately, see 
the corresponding suppression factor 
$ \exp [ -  2 \pi e^2\,Z_1 Z_2/ ( \hbar v_{12}'')]$
in  Eq.~(\ref{d3Q}).

\acknowledgements
This work was supported by the Australian Research Council.
The authors are grateful to V.G.Zelevinsky 
who kindly introduced the problem, 
and to H.Hora and J.S.M.Ginges for useful discussions.
One of us (B.L.) acknowledges the hospitality of 
the staff of the School of Physics of UNSW, where this work was
completed.

\begin{figure}
\caption{ \label{Rfactor}
The relative probability of the 
DD fusion initiated by TEC.
The ratio ${\cal R}$, defined in (\ref{F}),
is presented versus the projectile energy.
Target deuterons are implanted in an 
environment containing gold atoms
which are responsible for rescattering,
the geometry of collisions coincides with
the one  shown in Fig.~\ref{dd}.
Calculations are based on Eq.~(\ref{RR}) with the elastic cross
section from Fig.~\ref{cross}. 
The screening potential: 
thick solid line - no screening $U=0$,
thin solid line - $U=10$ eV,
dashed line - $U=20$ eV, 
dashed-dotted line - $U=30$ eV, 
dotted line - the strongest screening $U=50$ eV. }
\end{figure}

\end{document}